\documentclass[conference]{IEEEtran}
\IEEEoverridecommandlockouts
\usepackage{cite}
\usepackage{amsmath,amssymb,amsfonts}
\usepackage{algorithmic}
\usepackage{graphicx}
\usepackage{textcomp}
\usepackage{xcolor}
\usepackage{multirow}
\usepackage{multicol}
\usepackage{booktabs}
\usepackage{makecell}
\def\BibTeX{{\rm B\kern-.05em{\sc i\kern-.025em b}\kern-.08em
    T\kern-.1667em\lower.7ex\hbox{E}\kern-.125emX}}

\begin{document}

\title{Delving into Parameter-Efficient Fine-Tuning in Code Change Learning: An Empirical Study}

% \and
% \and
% \and
% \and
% \and

\author{
\IEEEauthorblockN{Shuo Liu\IEEEauthorrefmark{2},
Jacky Keung\IEEEauthorrefmark{2}, Zhen Yang\IEEEauthorrefmark{3}\IEEEauthorrefmark{1}\thanks{*Corresponding author.},
Fang Liu\IEEEauthorrefmark{4}\IEEEauthorrefmark{1}, Qilin Zhou\IEEEauthorrefmark{2},
and Yihan Liao\IEEEauthorrefmark{2}}
\IEEEauthorblockA{
\IEEEauthorrefmark{2}Department of Computer Science, City University of Hong Kong, Hong Kong, China, \\ \{sliu273-c, qilin.zhou,yihanliao4-c\}@my.cityu.edu.hk, Jacky.Keung@cityu.edu.hk\\
\IEEEauthorrefmark{3} School of Computer Science and Technology, Shandong University, Qingdao, China, zhenyang@sdu.edu.cn\\
\IEEEauthorrefmark{4}School of Computer Science and Engineering, Beihang University, Beijing, China, fangliu@buaa.edu.cn}
}

\maketitle

\begin{abstract}

Compared to Full-Model Fine-Tuning (FMFT), Parameter Efficient Fine-Tuning (PEFT) has demonstrated superior performance and lower computational overhead in several code understanding tasks, such as code summarization and code search. This advantage can be attributed to PEFT's ability to alleviate the catastrophic forgetting issue of Pre-trained Language Models (PLMs) by updating only a small number of parameters. As a result, PEFT effectively harnesses the pre-trained general-purpose knowledge for downstream tasks. However, existing studies primarily involve static code comprehension, aligning with the pre-training paradigm of recent PLMs and facilitating knowledge transfer, but they do not account for dynamic code changes. Thus, it remains unclear whether PEFT outperforms FMFT in task-specific adaptation for code-change-related tasks. To address this question, we examine two prevalent PEFT methods, namely Adapter Tuning (AT) and Low-Rank Adaptation (LoRA), and compare their performance with FMFT on five popular PLMs. Specifically, we evaluate their performance on two widely-studied code-change-related tasks: Just-In-Time Defect Prediction (JIT-DP) and Commit Message Generation (CMG). The results demonstrate that both AT and LoRA achieve state-of-the-art (SOTA) results in JIT-DP and exhibit comparable performances in CMG when compared to FMFT and other SOTA approaches. Furthermore, AT and LoRA exhibit superiority in cross-lingual and low-resource scenarios. We also conduct three probing tasks to explain the efficacy of PEFT techniques on JIT-DP and CMG tasks from both static and dynamic perspectives. The study indicates that PEFT, particularly through the use of AT and LoRA, offers promising advantages in code-change-related tasks, surpassing FMFT in certain aspects. This research contributes to a deeper understanding of the capabilities of PEFT in leveraging pre-trained PLMs for dynamic code changes. 
% The replication package is available at https://github.com/ishuoliu/PEFT4CC.
\end{abstract}

\begin{IEEEkeywords}
Pre-trained Language Models, Adapter Tuning, Low-Rank Adaptation, Code Change
\end{IEEEkeywords}

\section{Introduction}

Pre-trained Language Models (PLMs), e.g., CodeBERT \cite{feng-etal-2020-codebert} and CodeT5 \cite{wang-etal-2021-codet5},  have been extensively utilized in various software engineering tasks, including code generation \cite{lu2021codexglue, guo-etal-2022-unixcoder} and code comprehension \cite{karmakar2021pre, DBLP:conf/iclr/GuoRLFT0ZDSFTDC21}, and have achieved notable advancements \cite{xu2022survey}. These models follow a prevalent paradigm where they are initially pre-trained on large-scale monolingual corpora using Masked Language Modeling (MLM) or Next Token Prediction (NTP) \cite{DBLP:conf/naacl/DevlinCLT19}, and subsequently fine-tuned for specific downstream tasks. Nonetheless, Full-Model Fine-Tuning (FMFT) prohibitively relies on enormous computational resources, rendering it unsuitable for all users. Therefore, a series of Parameter-Efficient Fine-Tuning (PEFT) methods \cite{houlsby2019parameter, hu2022lora, li-liang-2021-prefix, lester-etal-2021-power}, updating only 1\%-5\% of the whole parameters while keeping the others frozen, have been proposed in recent years and gained much attention \cite{ding2023parameter, fu2023effectiveness}. 

\begin{figure}[t]
	\centering
        \setlength{\abovecaptionskip}{0.cm}
        
		\includegraphics[width=0.5\textwidth]{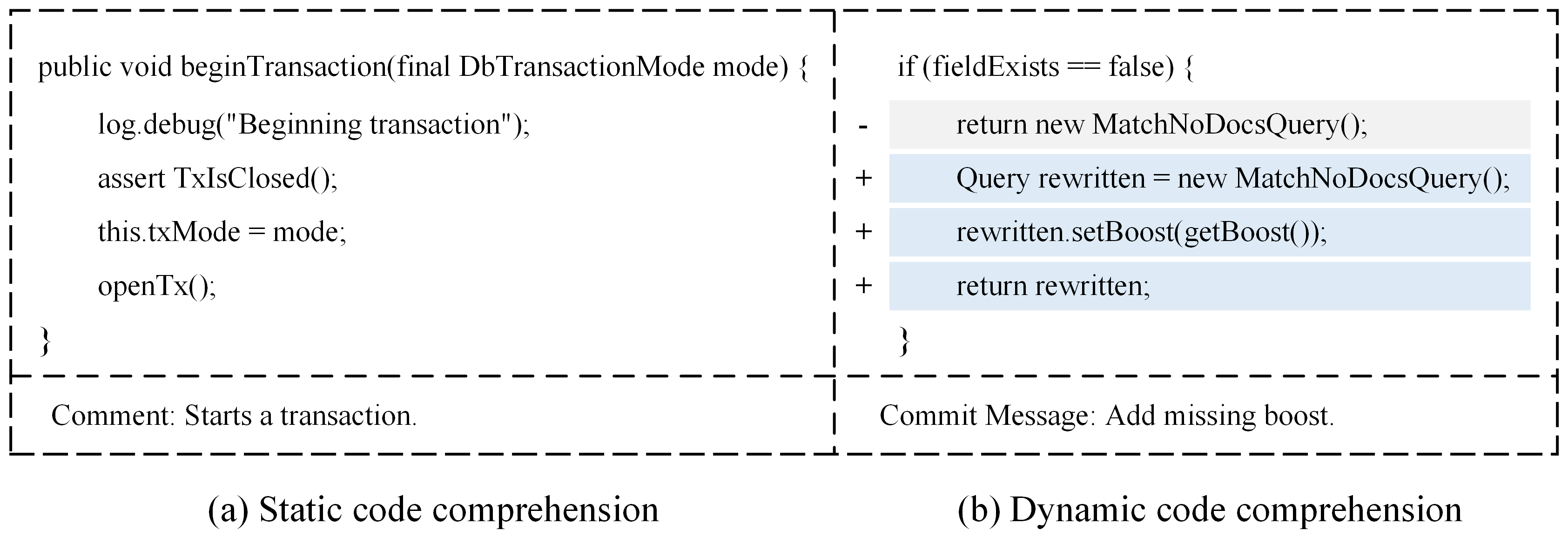}
	  \caption{Examples of static and dynamic code comprehension.}\label{code_change}
\vspace{-0.5cm}
\end{figure}

Previous studies \cite{saberi2023utilization, wang2023one} revealed the superiority of PEFT against FMFT in code search, code clone, and code summarization tasks. Because PEFT greatly alleviates the catastrophic forgetting problem, it can efficiently harness the pre-trained knowledge to the downstream tasks \cite{goel2022cross}. However, it is noted that most of the investigation objects of the above literature are static code comprehension tasks, excluding dynamic code changes, which is consistent with the paradigm that PLMs learned during their pre-training phase. As such, we conjecture adapting PLMs to static code comprehension tasks inherently needs fewer parameter tuning. However, for downstream tasks concerning code changes, both before and after-change code snippets are involved, enlarging the transferring gap between the pre-training and adaptation phases \cite{lin2023cct5,wang2022no}. Figure \ref{code_change} shows examples of static and dynamic code comprehension, where the former mainly relates to a single status of a code comment pair, while the latter contains a code \textit{diff} that compares program statuses of old and new versions line by line, described by a commit message recording the intent of changes. 
% It can be observed that static code semantics are mainly related to code functions, while code changes, which contains a code \textit{diff} that compares two programs line by line and the corresponding commit message, describe the intent of the change. 
% Due to their different forms
Considering their discrepancies above, whether PEFT still outperforms FMFT in code-change-related tasks remains an open question. 

In this paper, we explore the effect of PEFT on code change learning. We choose two mainstream PEFT methods for experiments, namely Adapter Tuning (AT) \cite{he-etal-2021-effectiveness, houlsby2019parameter, pfeiffer-etal-2020-adapterhub} and Low-Rank Adaptation (LoRA) \cite{hu2022lora}. AT introduces adapter modules, each of which is a bottleneck structure, and inserts them between layers of PLMs. In the fine-tuning stage, the parameters of the original PLMs are fixed, and only adapter modules are adjusted. LoRA injects trainable rank decomposition matrices into PLMs to substitute the original pre-trained weights, which can amplify some hidden features encoded during pre-training. To evaluate their performance on code change learning, we conduct experiments for Just-In-Time Defect Prediction (JIT-DP) \cite{zeng2021deep} and Commit Message Generation (CMG) \cite{cortescoy2014auto}. The former is a classification task aiming to predict whether a code change is defect-prone or not, and the latter is a generation task targeting to automatically generate a commit message given changed code snippets. 

Specifically, we first compare the performance of AT and LoRA with FMFT and state-of-the-art approaches. Furthermore, to evaluate how well AT and LoRA transfer knowledge and generalize across languages, we explore their capabilities in the cross-lingual scenario, where PLMs are fine-tuned in one programming language but are tested in another. Besides, it is widely acknowledged that the performance of FMFT is highly dependent on the amount of available downstream data. In order to understand how well PEFT performs in scenarios where data availability is limited, we conducted experiments to examine its effectiveness in handling situations of data scarcity, where the fine-tuned datasets are scaled down, but the testing datasets remain the same. We also utilize three probing tasks, namely Invalid Type Detection (TYP), Code Change Match (CCM), and Line Type Prediction (LTP), to investigate the encoded code semantics, expecting to make an explanation for the performance of PEFT. TYP is related to the static code semantics, while CCM and LTP are associated with the dynamic code semantics from global and local perspectives, respectively. Our experimental results show that, in the JIT-DP task, AT and LoRA obtain improvements of 8.39\% and 9.87\% in terms of F1 when compared with the State-Of-The-Art (SOTA) baseline, and in the CMG task, they perform a little weakly. 
% The average performance of FMFT, AT, and LoRA are 22.16, 21.60, and 19.85 in terms of BLEU. 
However, considering the less training time and memory consumption, PEFT techniques are still practical and acceptable. Besides, even in the cross-lingual and low-resource scenarios, they also exhibit effectiveness and superiority. Three probing tasks concerning both static and dynamic code semantics explain the efficacy of PEFT techniques and FMFT. 
% The probing tasks indicate that the improvements obtained on the JIT-DP task can be attributed to the better comprehension that AT and LoRA have about the code-chang-related semantics.

\begin{figure*}[t]
	\centering
        \setlength{\abovecaptionskip}{0.cm}
		\includegraphics[width=\textwidth]{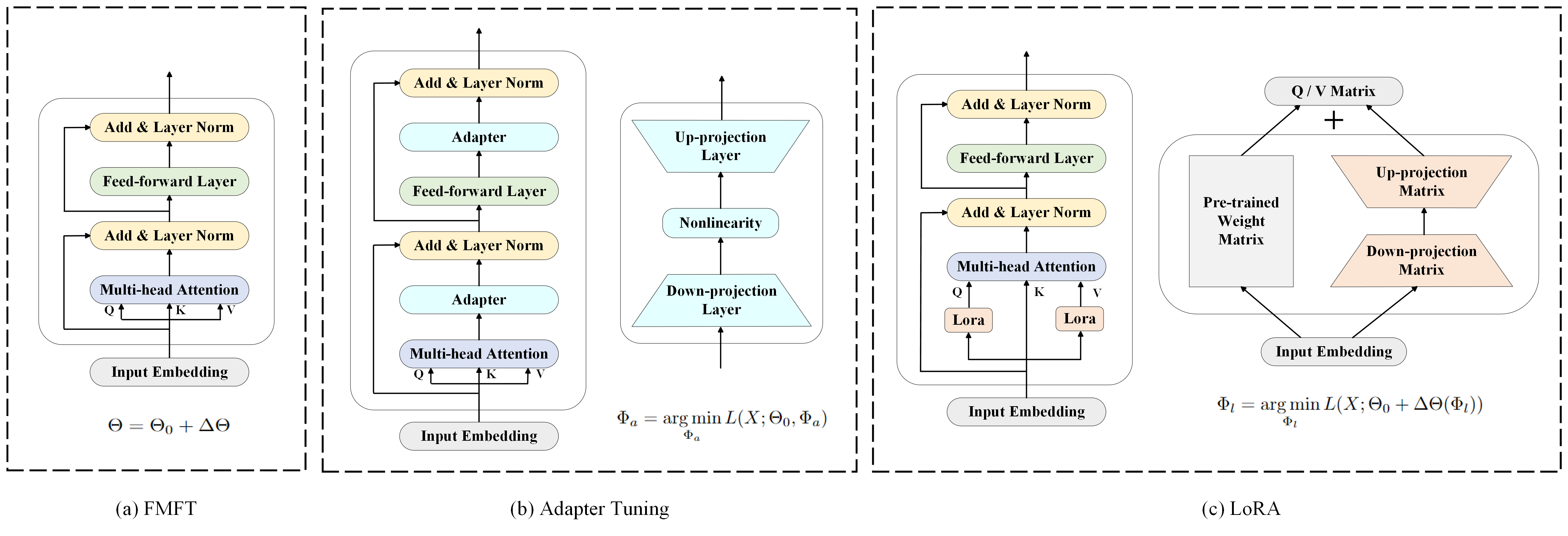}
	  \caption{Illustrations on fine-tuning, adapter tuning, and low-rank adaptation.}\label{illustrations}
\vspace{-0.5cm}
\end{figure*}

The major contributions of this paper are as follows:

\begin{itemize}
\item [1.] To the best of our knowledge, this paper serves as the first study to explore the performance of PEFT on code-change-related tasks.
\item [2.] We conduct extensive experiments, involving two PEFT methods, five PLMs, and two specific code-change-related tasks. We also explore their performance in the cross-lingual and low-resource scenarios.
\item [3.] We propose two code-change-related probing tasks to investigate the encoded dynamic code semantics and construct corresponding datasets.
% \item[4.] We open-source the replication package of this work for future research and application\footnote{https://zenodo.org/records/10053278}.
\end{itemize}

% Pre-trained Language Models have been widely applied to downstream tasks in the code domain, and have achieved considerable success. The use of PLMs follows the pretrain-then-finetune paradigm. PLMs are first pre-trained from scratch on massive code-related data, to obtain general comprehension of code. There have been several representative pre-trained PLMs like CodeBERT \cite{feng-etal-2020-codebert} and CodeT5 \cite{wang-etal-2021-codet5}. And then PLMs are fine-tuned in downstream tasks to transfer the pre-learned knowledge, which can mitigate the discrepancy between the pre-training and specific downstream tasks. 

% However, for code-change-related tasks, the discrepancy is huge. 

% One of the main drawbacks of fine-tuning is that it prohibitively relies on computational resources. PLMs usually have parameters on the magnitude of millions even billions, which makes fine-tuning not only a large consumption of memory and time, but also impractical for everyone \cite{lialin2023scaling}. Alternatively, parameter-efficient fine-tuning appears, which aims at updating a small set of added or selected parameters while keeping the others frozen, and has achieved surprisingly good performance in the NLP field \cite{ding2023parameter, fu2023effectiveness}.

\section{Background}

\subsection{Full-Model Fine-Tuning}

Pre-training and fine-tuning have become the dominant paradigm for applying PLMs to specific downstream tasks, and have shown promising results on many code-related tasks \cite{kang2023large, lemieux2023codamosa, xia2023automated, yang2024improving}. In the pre-training stage, PLMs are trained from scratch to learn general-purpose representations, and in the fine-tuning stage, PLMs utilize the pre-trained parameters as initialization to adapt them to downstream tasks. Formally, given a downstream task-specific dataset $X$ and corresponding labels $Y$, FMFT initializes PLMs to pre-trained weights $\Theta_0$ and updates them to $\Theta$ in a supervised manner, which can be formulated as:
\begin{equation}
\Theta = \Theta_0 + \Delta\Theta = \mathop{\arg\min}_{\Theta} L(X; \Theta)
\end{equation}
where $L$ is a loss function and $\Delta\Theta$ represents the adjustment of parameters. Considering that PLMs are Transformer-based architecture \cite{vaswani2017attention}, FMFT is illustrated in Figure \ref{illustrations}(a). The parameters of the core block of the Transformer, including multi-head attention, feed-forward layer, and layer normalization, as well as the projection matrices $W_Q$, $W_K$, and $W_V$, all are part of $\Theta$ and are trainable in FMFT.

\subsection{Adapter Tuning}

Adapter tuning is a representative method of PEFT, which adds extra compact modules (adapters) to every transformer layer \cite{he-etal-2021-effectiveness, houlsby2019parameter, pfeiffer-etal-2020-adapterhub}. Figure \ref{illustrations}(b) shows the standard architecture of adapter tuning \cite{houlsby2019parameter}, where adapter modules are added twice after the multi-head attention and the feed-forward layer. Supposing that adapter modules are fed $d$-dimensional features, they first project features into dimension $m$, $m \ll d$, then apply a nonlinearity, and finally project back to dimension $d$. The two projection layers construct a bottleneck structure. Specifically, adapter modules are initialized randomly, and the target of adapter tuning is to train the added modules and corresponding following layer normalizations. Its tuning objective can be defined as:
\begin{equation}
\Phi_a = \mathop{\arg\min}_{\Phi_a} L(X; \Theta_0, \Phi_a)
\end{equation}
where $\Phi_a$ is the adapter parameter. $X$ and $L$ are task-specific dataset and loss function respectively. PLMs' pre-trained weights $\Theta_0$ are frozen here. During adapter tuning, around 3\% - 5\% of the whole parameters are trainable, which is considerably fewer than fine-tuning.

\subsection{Low-Rank Adaptation (LoRA)}

Different from adapter tuning that adds extra trainable parameters, LoRA utilizes the idea of reparameterization to learn the parameters' updation by low-rank matrices \cite{lialin2023scaling}. As illustrated in Figure \ref{illustrations}(c), when applying LoRA to the transformer architecture, it is limited to only adapting the attention modules, and specifically, adapting weight matrices $W_Q$ and $W_V$ \cite{hu2022lora}. For a pre-trained weight matrix $W \in \mathbb{R}^{d \times d}$, LoRA decomposes its updation into two low-rank matrices that satisfy $\Delta W = BA$, where $B \in \mathbb{R}^{d \times r}$, $A \in \mathbb{R}^{r \times d}$, $r \ll d$. Supposing $\Delta \Theta$ is a task-specific parameter modification, it can be further encoded by the LoRA parameter $\Phi_l$ as $\Delta \Theta(\Phi_l)$. The tuning objective is then transferred to optimize $\Phi_l$ as:
\begin{equation}
\Phi_l = \mathop{\arg\min}_{\Phi_l} L(X; \Theta_0 + \Delta \Theta(\Phi_l))
\end{equation}
where $X$ and $L$ are the dataset and loss function of downstream tasks. $\Theta_0$ is also fixed here. Due to LoRA being only inserted into weight matrices $W_Q$ and $W_V$, as well as the dimension $r \ll d$, 
% the trainable parameters of LoRA can be as small as 1\% of the whole parameters. 
the trainable parameters of LoRA can be reduced to 1\%.

\section{Experimental Setup}

\subsection{Research Questions}

Although previous studies have shown the effectiveness of PEFT in code search, code clone, and code summarization \cite{saberi2023utilization, wang2023one}, these tasks are limited in the comprehension of static code semantics. For code-change-related tasks that require the understanding of dynamic code semantics, there is still an open question about the efficacy of PEFT. To address the above issue, we investigate the following research questions about adapter tuning and LoRA, exploring their effects and advantages.

\textbf{RQ1: How do adapter tuning and LoRA perform compared with FMFT and SOTA approaches?}
Previous studies have shown the effectiveness of PEFT methods in static code comprehension tasks, we investigate whether they can perform consistent advantages in code-change-related tasks. Specifically, we compare the performance of FMFT, AT, and LoRA on two widely studied code-change-related tasks, Just-In-Time Defect Prediction (JIT-DP) and Commit Message Generation (CMG). The former is a classification task, and the latter is a generation task. For JIT-DP, we apply AT and LoRA to five popular PLMs, namely CodeBERT, GraphCodeBERT, PLBART, UniXcoder, and CodeT5, to conduct comprehensive comparisons and explore the generalization abilities of PEFT on diverse PLMs. Besides, various task-specific SOTA approaches are also involved for comparison \cite{pornprasit2021jitline, hoang2020cc2vec, lin2023cct5, ni2022best}. For CMG, We apply the three methods on CodeT5 because it is the backbone model of current SOTA approaches \cite{li2022automating, lin2023cct5}. 

\textbf{RQ2: How capable are adapter tuning and LoRA in the cross-lingual scenario compared with FMFT?} 
In the cross-lingual scenario, pre-trained PLMs are fine-tuned in one Programming Language (PL) and are evaluated in another. It is a common but resultful approach to relieve the problem of data scarcity \cite{zhu2022xlcost}. To explore the performance of adapter tuning and LoRA in the cross-lingual scenario, we carry out experiments with CodeT5 on the CMG task, which involves five PLs. For each PL, we attempt FMFT, AT, and LoRA for adaptation, then evaluate them on each of the other PLs.
% in the CMG task, which contains five PLs and has corresponding fine-tuned models, we evaluate the fine-tuned models on all PLs respectively. 

\textbf{RQ3: How capable are adapter tuning and LoRA in the low-resource scenario compared with FMFT?} 
The performance of FMFT prohibitively relies on the scale of downstream data \cite{gu-etal-2022-ppt, lester-etal-2021-power, zhang2022differentiable}. However, large-scale datasets are usually rare. In addition, fine-tuning in the low-resource scenario means that PLMs can learn task-specific knowledge quickly, which meets practitioners' usual expectations. Considering that the CMG task contains hundreds of thousands of data, we investigate the performance of adapter tuning and LoRA when randomly sampling 1,000, 5,000, and 10,000 training data for each PL. We still evaluate the models on the original validation set and testing set. 

\textbf{RQ4: What kind of knowledge is encoded by adapter tuning and LoRA?} 
To explore what benefits adapter tuning and LoRA can bring about, we utilize probing tasks, which have been extensively used in the NLP field \cite{tenney-etal-2019-bert, tenney2018what}, to understand the encoded linguistic information. We employ three probing tasks, invalid type detection for static semantic-level information \cite{karmakar2021pre}, as well as code change match and line type prediction, which are proposed by us to probe the dynamic semantic-level information from the global and local perspectives, respectively. The three probing tasks are introduced in detail in section \ref{tasks datasets}. We mainly investigate what kind of knowledge is encoded by adapter tuning and LoRA, as well as making comparisons with FMFT, thereby providing explanations for their performance.

\subsection{Tasks and Datasets} \label{tasks datasets}

\begin{table}[t]
\caption{Dataset Statistics.} \label{dataset statistics}
\centering
\begin{tabular}{lllll}
\hline
\textbf{Tasks} & \textbf{Datasets} & \textbf{Training} & \textbf{Validation} & \textbf{Test} \\
\hline
\textbf{JIT-DP} & JIT-Defects4J & 16,374 & 5,465 & 5,480 \\
\hline
\multirow{5}{*}{\textbf{CMG}} & Java & 160,018 & 19,825 & 20,159 \\
& C\# & 149,907 & 18,688 & 18,702 \\
& C++ & 160,948 & 20,000 & 20,141 \\
& Python & 206,777 & 25,912 & 25,837 \\
& JavaScript & 197,529 & 24,899 & 24,773 \\
\hline
\multirow{3}{*}{\textbf{\thead[l]{Probing\\Tasks}}} & TYP & 600 & 200 & 200 \\
& CCM & 600 & 200 & 200 \\
& LTP & 600 & 200 & 200 \\
\hline
\end{tabular}
\label{datasets}
\vspace{-0.5cm}
\end{table}

\subsubsection{Just-In-Time Defect Prediction}
The change of code may damage software quality, so it is crucial to discover defects as early as possible \cite{zeng2021deep}. JIT-DP aims to identify defective code changes when they are just committed, and return judgements. It can be seen as a binary classification task for the outputs are two classes, namely defective or not.

The experimental dataset we used in JIT-DP is JIT-Defects4J \cite{ni2022best}, which is extended from LLTC4J \cite{herbold2022fine} with extra buggy commits. In addition, for each code change, JIT-Defects4J extracts the 14 change-level Expert Features (EF) proposed by Kamei et al. \cite{kamei2012large}, which can measure the code change from five dimensions, i.e. diffusion, size, purpose, history, and experience. Following previous studies \cite{lin2023cct5, ni2022best}, we evaluate different methods in two settings, one for directly using the learned code change representations to predict, and another for incorporating the expert features.

\subsubsection{Commit Message Generation}

Commit message summarizes the intent and content of a code change, which is helpful for developers to understand programs quickly in software maintenance \cite{cortescoy2014auto}. However, writing high-quality commit messages is time-consuming, and thus many are left empty \cite{dyer2013boa}. Therefore, taking changed codes as inputs, the CMG task aims to generate commit messages automatically and has become a hot topic in the software engineering domain.

In the CMG task, we choose the Multi-language Commit Message Dataset (MCMD) for experiments \cite{tao2021evaluation}. MCMD contains five PLs, including Java, C\#, C++, Python, and Javascript. For each PL, it collects the top 100 starred projects from GitHub and randomly retains 450,000 commits. Furthermore, following the operation of Shi et al. \cite{shi-etal-2022-race}, noisy data containing multiple files or files that cannot be parsed are filtered out. The statistics of MCMD are shown in Table \ref{dataset statistics}.

\subsubsection{Probing Tasks}

\begin{figure*}[t]
	\centering
        \setlength{\abovecaptionskip}{0.cm}
		\includegraphics[width=\textwidth]{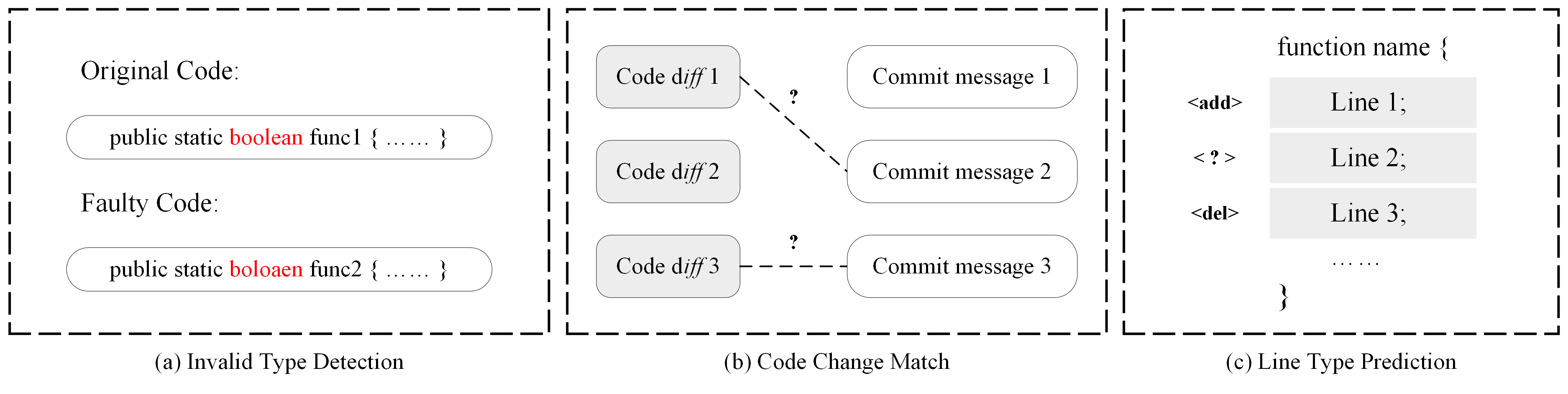}
	  \caption{Illustrations of probing tasks.}\label{probe_illustration}
\vspace{-0.5cm}
\end{figure*}

We employ three probing tasks that are related to code properties from different aspects. For their datasets, we split them into the training set, validation set, and testing set by the ratio of 6/2/2 uniformly.

\textbf{Invalid Type Detection (TYP)}, proposed by Karmakar et al. \cite{karmakar2021pre}, aims to measure the static code semantics understood by PLMs. As Figure \ref{probe_illustration}(a) shows, it falsifies the data types of some code snippets deliberately, and expects PLMs to distinguish the invalid samples from the others. We introduce the probing task here to explore whether PLMs still encode static code semantics when they are fine-tuned in code-change-related tasks. We also adopt the dataset containing 1,000 samples from Karmakar et al. \cite{karmakar2021pre}. It is gathered from 50K-C \cite{martins2018dataset}, a dataset of compilable Java projects, and balances the valid and invalid classes. 

\textbf{Code Change Match (CCM)}, probing the dynamic semantic-level information, is proposed by us. Different from TYP, designed for code semantics in one code snippet, CCM takes a pair of changed codes and a commit message as inputs, inferring whether a commit message corresponds to a given code change. It is shown in Figure \ref{probe_illustration}(b). If the judgment is correct, it means PLMs can identify the code change semantics from a global perspective. We construct a dataset for CCM from FIRA \cite{dong2022fira}, which collects commits from the top 1,000 popular Java projects in GitHub. Similar to the dataset of TYP, we keep 500 correct matches and randomly produce 500 false matches.

\textbf{Line Type Prediction (LTP)}, proposed by us as well, is also to determine whether PLMs can encode dynamic semantic-level information. As code changes are composed of added, kept, and deleted lines, LTP aims to predict the line type when given the surrounding context, as the illustration in Figure \ref{probe_illustration}(c).
% As the basis of code changes is the added or deleted lines, when given a masked line type and its corresponding context, LTP should predict the masked type.
Considering LTP is designed for predicting specific line types, it measures PLMs' ability to comprehend the semantics of code changes from a local perspective. We also build the dataset for LTP from FIRA \cite{dong2022fira}. For each commit, 
% FIRA divides its lines into three categories: added, deleted, and kept.
to measure the perception of code changes, we only retain the added and deleted lines, then randomly mask a line type for prediction. The dataset also incorporates 1,000 samples, and Table \ref{dataset statistics} demonstrates its statistics.

\subsection{Pre-trained Language Models and Baselines}
There are five PLMs that we use to evaluate the performance of FMFT, adapter tuning, and LoRA. In the JIT-DP task, we conduct experiments on all of the five PLMs, where only encoders are used for the purpose of code change representation. Besides, we choose CodeT5 as the backbone model for the CMG task because it is used by the SOTA baselines \cite{li2022automating, lin2023cct5}. The details of the PLMs are described in the following:
\begin{itemize}
\item \textbf{CodeBERT} \cite{feng-etal-2020-codebert}: It is an encoder-only PLM with 125M parameters, pre-trained on CodeSearchNet \cite{husain2019codesearchnet} which contains 2M bimodal data.
\end{itemize}
\begin{itemize}
\item \textbf{GraphCodeBERT} \cite{DBLP:conf/iclr/GuoRLFT0ZDSFTDC21}: It has the same architecture and parameter size as CodeBERT, while pre-trained with edge prediction and node alignment to learn from data flow.
\end{itemize}
\begin{itemize}
\item \textbf{PLBART} \cite{ahmad-etal-2021-unified}: It is pre-trained on Java and Python datasets, as well as natural language descriptions. It is an encoder-decoder architecture that has 140M parameters.
\end{itemize}
\begin{itemize}
\item \textbf{UniXcoder} \cite{guo-etal-2022-unixcoder}: It is a unified encoder-decoder PLM with 125M parameters, and can be flexibly modified to encoder-only or decoder-only architecture.
\end{itemize}
\begin{itemize}
\item \textbf{CodeT5} \cite{wang-etal-2021-codet5}: It contains 220M parameters, and is a representative PLM of encoder-decoder architecture for its well performance in generation tasks.
\end{itemize}

We also compare the performance of FMFT, adapter tuning, and LoRA with the following baselines. 

\begin{itemize}
\item \textbf{JITLine} \cite{pornprasit2021jitline}: It is a baseline of JIT-DP, which represents code changes by bag-of-tokens features and constructs classifiers like SVM to predict defective commits.
\end{itemize}
\begin{itemize}
\item \textbf{JITFine} \cite{ni2022best}: It uses CodeBERT as an encoder, and incorporates the encoded code change representations with extra expert features, achieving a substantial improvement in the JIT-DP task.
\end{itemize}
\begin{itemize}
\item \textbf{CC2Vec} \cite{hoang2020cc2vec}: It is a baseline of JIT-DP, which utilizes a hierarchical attention network and emphasizes modeling the correlation between removed codes and added codes.
\end{itemize}
\begin{itemize}
\item \textbf{CodeReviewer} \cite{li2022automating}: It is proposed for code review activities, but can be extended to the CMG task. It proposes four pre-training tasks to improve CodeT5's \cite{wang-etal-2021-codet5} capacity for understanding code changes.
\end{itemize}
\begin{itemize}
\item \textbf{CCT5} \cite{lin2023cct5}: It is the state-of-the-art baseline for both JIT-DP and CMG. It also pre-trains CodeT5 \cite{wang-etal-2021-codet5} with code-change-related corpus and objectives.
\end{itemize}

\subsection{Evaluation Metrics}

\subsubsection{Just-In-Time Defect Prediction}
Following prior works \cite{lin2023cct5, ni2022best}, we use the F1-score and the Area Under the receiver operating characteristic Curve (AUC) as the evaluation metrics. F1-score combines the precision and recall scores of a model, and AUC reflects the distinguishing ability of models between positive and negative classes. In the JIT-DP task, we treat defect-prone code changes as positive samples.

\subsubsection{Commit Message Generation}
For the CMG task, we evaluate the generated commit messages using three metrics: BLEU \cite{papineni2002bleu}, Meteor \cite{banerjee2005meteor}, and Rouge-L \cite{lin2004rouge}. BLEU calculates the n-gram precision between the generated text and ground truth text. Meteor takes into account the precision, recall, and fluency of the generated text. Rouge-L focuses on the longest common subsequence so that it evaluates more about the word order. Specifically, for the BLEU metric, we choose a smoothed BLEU-4 score for evaluations in this paper. 

% which is defined as:
% \begin{equation}
% BP = 
% \begin{cases}
% 1 & \text{if} \ c > r \\
% e^{1-r/c} & \text{if} \ c \leq r
% \end{cases}
% \end{equation}

% \begin{equation}
% BLEU = BP \cdot \exp(\sum_{n=1}^N w_n \log p_n)
% \end{equation}
% where $p_n$ is the n-gram precision and $w_n$ is a uniform weight $1/N$. BP is a brevity penalty to avoid generating short sentences. $c$ is the length of the generated text, and $r$ is the length of the ground truth text.

\subsubsection{Probing Tasks}
Due to probing tasks being classification tasks, we use accuracy as the evaluation metric, which measures the correctness of predictions.

\subsection{Implementation Details} \label{implementation details}

The experiments were conducted on a Ubuntu GPU server with two RTX 3090 24GB GPUs. Our code is implemented by the deep learning framework PyTorch\footnote{https://pytorch.org/}. The PLMs we used are from Huggingface\footnote{https://huggingface.co/models}, and we keep their default hyperparameter settings. For the adapter tuning, we implement it with the OpenDelta Library\footnote{https://opendelta.readthedocs.io/en/latest/}, and we set the intermediate dimension $m$ to 128. LoRA is implemented by the PEFT Library\footnote{https://huggingface.co/docs/peft/index}, and its dimension $r$ is 8. In the JIT-DP task, we adjust the learning rate in the range of \{1e-3, 5e-4, 1e-4, 5e-5\} for all PLMs. In the CMG task, we set the learning rate of CodeT5 to 5e-4 and keep it consistent across FMFT, adapter tuning, and LoRA. Besides, we set the maximum training epoch to 10, and adopt early stopping with the patience of 5. Due to the performance of baselines on our evaluation datasets having been investigated in previous studies, we reuse the performance reported by Ni et al. \cite{ni2022best} and Lin et al. \cite{lin2023cct5}.

\section{Experimental Results}

\subsection{RQ1: Quantitative Evaluation}

\begin{table}[t]
\setlength\tabcolsep{5.5pt}
\caption{Evaluation results of JIT-DP on the JIT-Defects4J dataset.} \label{JIT-DP main}
\centering
\begin{tabular}{cccccc}
\hline
\multicolumn{2}{c}{\multirow{2}*{\textbf{Models}}} & \multicolumn{2}{c}{\textbf{w/o EF}} & \multicolumn{2}{c}{\textbf{w/ EF}} \\
\cline{3-6}
& & F1 & AUC & F1 & AUC \\
\hline
\multirow{3}{*}{\textbf{\thead{End-to-end}}} & JITLine & 0.261 & 0.802 & - & - \\
& JITFine & 0.375 & 0.856 & 0.431 & 0.881 \\
& CC2Vec & 0.248 & 0.791 & - & - \\
\hline
\textbf{Pre-trained} & CCT5 & \textbf{0.451} & \textbf{0.871} & 0.472 & 0.882 \\
\hline
\multirow{6}{*}{\textbf{\thead{FMFT}}} & CodeBERT & 0.382 & 0.849 & 0.419 & 0.867 \\
& GraphCodeBERT & 0.390 & 0.869 & 0.361 & 0.861 \\
& PLBART & 0.329 & 0.834 & 0.412 & 0.869 \\
& UniXcoder & 0.401 & 0.847 & 0.480 & 0.889 \\
& CodeT5 & 0.398 & 0.859 & 0.433 & 0.878 \\
\cline{2-6}
& Avg. & 0.380 & 0.852 & 0.421 & 0.873 \\
\hline
\multirow{6}{*}{\textbf{\thead{LoRA}}} & CodeBERT & 0.380 & 0.860 & 0.512 & 0.899 \\
& GraphCodeBERT & 0.379 & 0.869 & 0.523 & 0.898 \\
& PLBART & 0.393 & 0.862 & 0.512 & 0.890 \\
& UniXcoder & 0.388 & 0.852 & 0.527 & 0.907 \\
& CodeT5 & 0.376 & 0.867 & 0.519 & 0.895 \\
\cline{2-6}
& Avg. & 0.383 & 0.862 & \textbf{0.519} & \textbf{0.898} \\
\hline
\multirow{6}{*}{\textbf{\thead{Adapter\\Tuning}}} & CodeBERT & 0.408 & 0.864 & 0.502 & 0.894 \\
& GraphCodeBERT & 0.377 & 0.866 & 0.525 & 0.897 \\
& PLBART & 0.377 & 0.858 & 0.494 & 0.890 \\
& UniXcoder & 0.402 & \textbf{0.871} & 0.515 & 0.901 \\
& CodeT5 & 0.404 & 0.866 & 0.522 & 0.892 \\
\cline{2-6}
& Avg. & 0.394 & 0.865 & 0.512 & 0.895 \\
\hline
\end{tabular}
\label{datasets}
\vspace{-0.5cm}
\end{table}

\subsubsection{Just-In-Time Defect Prediction}
We first evaluate the performance of adapter tuning and LoRA in the JIT-DP task. We employ five prevalent PLMs to exhibit comprehensive comparisons, including CodeBERT, GraphCodeBERT, PLBART, UniXcoder, and CodeT5. The latter three PLMs are encoder-decoder architecture, and we only use their encoders in the task. Following previous studies \cite{lin2023cct5, ni2022best}, we also conduct experiments on two settings. The first is a common setting (w/o EF), where encoded code change representations are extracted and then fed into a linear classifier, predicting defectiveness or not. The second setting (w/ EF) combines the encoded representations with extra expert features \cite{kamei2012large}, and the concatenated feature vectors are used to predict. Table \ref{JIT-DP main} shows the detailed evaluation results. Because they do not utilize expert features in their original papers, the performance of JITLine and CC2Vec in the second setting is omitted.

When comparing the performance of FMFT, adapter tuning, and LoRA, it can be noticed that no matter in which setting, the average performance of adapter tuning and LoRA is better than FMFT. Specifically, in the common setting, adapter tuning obtains an improvement of 3.58\% in terms of F1 and 1.57\% in terms of AUC, while LoRA increases by 0.84\% in terms of F1 and 1.22\% in terms of AUC. When incorporating extra expert features, the improvements become significant. Adapter tuning surpasses fine-tuning by 21.52\% and 2.52\% for F1 and AUC, respectively. LoRA increases 23.18\% and 2.86\% accordingly. The results indicate that adapter tuning and LoRA show their effectiveness in the JIT-DP task, especially when introducing extra expert features. A potential explanation is adapter tuning and LoRA with very few parameter tuning are more capable of learning straightforward features like expert features, thereby obtaining substantial improvement.
% Especially when introducing extra expert futures, they perform their superiority in encoding expert knowledge in PLMs. 
In addition, it can be observed that adapter tuning and LoRA consistently show their improvements in diverse PLMs, which manifests their generalization abilities.

When compared with other baselines, adapter tuning and LoRA achieve state-of-the-art results that reach 0.512 and 0.519 in terms of F1 in the setting with extra expert features. They surpass CCT5 \cite{lin2023cct5}, the current SOTA method, by 8.39\% and 9.87\%, respectively. Considering that CCT5 has to be pre-trained from scratch on code-change-related corpus and various objectives, it indicates the prominent advantage of adapter tuning and LoRA because they only need to fine-tune a very small portion of the parameters.

\begin{table*}[t]
\setlength\tabcolsep{3.25pt}
\caption{Evaluation results of CMG on the MCMD dataset. M is short for Meteor. R is short for Rouge-L. CodeT5, CodeT5-L, and CodeT5-A represent they utilize FMFT, Adapter Tuning, and LoRA, respectively.} \label{CMG main}
\centering
\scalebox{1}{
    \begin{tabular}{ccccccccccccccccccc}
    \hline
    \multirow{2}*{\textbf{Models}} & \multicolumn{3}{c}{\textbf{Java}} & \multicolumn{3}{c}{\textbf{C\#}} & \multicolumn{3}{c}{\textbf{C++}} & \multicolumn{3}{c}{\textbf{Python}} & \multicolumn{3}{c}{\textbf{JavaScript}} & \multicolumn{3}{c}{\textbf{Avg.}} \\
    \cline{2-19}
    & BLEU & M & R & BLEU & M & R & BLEU & M & R & BLEU & M & R & BLEU & M & R & BLEU & M & R \\
    \hline
    CodeReviewer & 18.47 & - & - & 20.36 & - & - & 15.94 & - & - & 17.65 & - & - & 19.84 & - & - & 18.45 & - & - \\
    CCT5 & 20.80 & - & - & \textbf{25.53} & - & - & 17.64 & - & - & \textbf{21.37} & - & - & 24.94 & - & - & 22.06 & - & - \\
    \hline
    CodeT5 & \textbf{24.34} & \textbf{15.28} & \textbf{32.15} & 22.11 & \textbf{13.96} & \textbf{28.93} & \textbf{18.78} & 13.38 & \textbf{25.70} & 20.55 & \textbf{14.97} & \textbf{28.96} & \textbf{25.03} & \textbf{17.30} & \textbf{32.99} & \textbf{22.16} & \textbf{14.98} & \textbf{29.75} \\
    CodeT5-L & 21.65 & 14.14 & 29.17 & 18.39 & 11.89 & 24.84 & 16.97 & 12.60 & 23.36 & 18.76 & 14.02 & 26.57 & 23.48 & 16.18 & 31.17 & 19.85 & 
    13.77 & 27.02 \\
    CodeT5-A & 23.95 & 15.25 & 31.60 & 20.69 & 13.36 & 27.49 & 18.33 & \textbf{13.39} & 25.31 & 20.12 & 14.75 & 28.44 & 24.92 & 17.07 & 32.75 & 21.60 & 14.76 & 29.12 \\
    \hline
    \end{tabular}
}
\label{datasets}
\vspace{-0.5cm}
\end{table*}

\begin{center}
\fcolorbox{black}{gray!10} 
{
\parbox{.9\linewidth}
{\textbf{Finding 1}: Compared with FMFT, Adapter tuning and LoRA show their effectiveness in the JIT-DP task, especially when incorporating extra expert features. They obtain state-of-the-art results that achieve improvements of 8.39\% and 9.87\% in terms of F1 sore compared with the SOTA approach, respectively.}
}
\end{center}

\subsubsection{Commit Message Generation}
We also evaluate adapter tuning and LoRA on a widely-studied code-change-related generation task, namely commit message generation. We employ CodeT5 as the backbone PLM following previous studies \cite{li2022automating, lin2023cct5}. Considering there are five sub-datasets containing diverse PLs, namely Java, C\#, C++, Python, and JavaScript, we fine-tune PLMs for each PL separately and calculate the average performance among the five PLs. The evaluation results are shown in Table \ref{CMG main}. Because CodeReviewer and CCT5 utilize BLEU to evaluate their models, their performance on the Meteor and Rouge-L is omitted here. 

\begin{figure}[t]
	\centering
        \setlength{\abovecaptionskip}{0.cm}
		\includegraphics[width=0.5\textwidth]{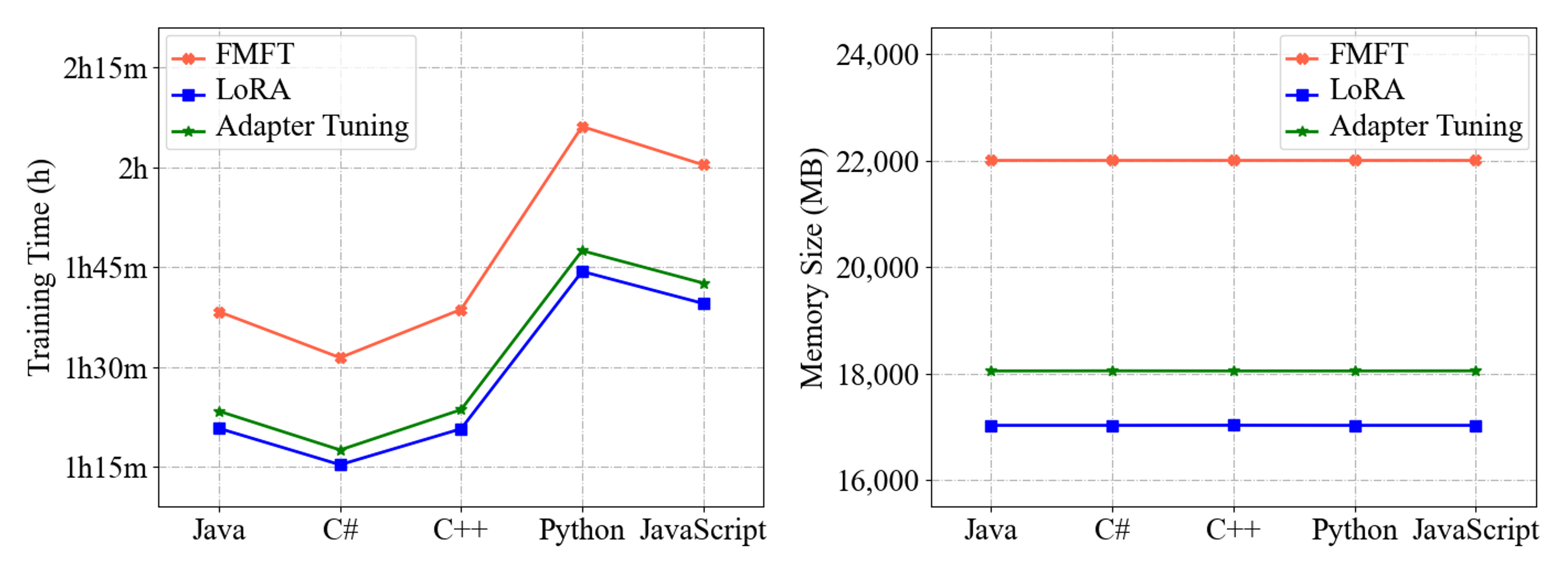}
	  \caption{Comparison of training time and memory size.}\label{plot_figure}
\vspace{-0.5cm}
\end{figure}

From Table \ref{CMG main}, we could observe mild degradations of adapter tuning and LoRA when comparing them with FMFT. The degradation is consistent on the five PLs. The average performance of adapter tuning and LoRA are 21.60 and 19.85 in terms of BLEU, both of which are lower than the FMFT performance of 22.16. We conjecture the reasons for such decrements mainly contain: \text{(i)} In the architecture of CodeT5, adapter tuning and LoRA only modify 4.10\% and 0.40\% of the full parameters, which is quite a small portion of parameters. \text{(ii)} Compared with classification tasks, e.g., the JIT-DP task, generative tasks, such as the CMG task, carry a high complexity and difficulty. As such, the CMG task needs more parameter tuning for knowledge adaptation.   
% Together with the complexity of the CMG task and the large-scale training dataset, adapter tuning and LoRA can not encode sufficient task-specific knowledge into PLMs. 
The performance of adapter tuning is better than LoRA in the CMG task, and we attribute it to the more adapted parameters that adapter tuning has. In addition to evaluating the performance, we analyze the training time and memory size of the three methods, and the results are illustrated in Figure \ref{plot_figure}. We report the average training time of each epoch and the occupied memory size on a GPU. The devices we use are elaborated in Section \ref{implementation details}. It can be observed that adapter tuning and LoRA can save approximately 15-20 minutes each epoch, and decrease memory consumption of about 4,000-5,000 MB. It indicates that they can achieve similar performance compared with FMFT, but in less training time and memory consumption. 

Although CodeReviewer and CCT5 are pre-trained on code change datasets, when compared with them, adapter tuning and LoRA still perform well. They both surpass CodeReviewer on the average performance. Besides, adapter tuning outperforms CCT5 on two sub-datasets, Java and C++. It is well known that pre-training on large-scale datasets is not practical for everyone, but adapter tuning and LoRA can obtain similar results by modifying only a small portion of the parameters.

\begin{center}
\fcolorbox{black}{gray!10} 
{
\parbox{.9\linewidth}
{\textbf{Finding 2}: In the CMG task, adapter tuning and LoRA can achieve similar performances compared with fine-tuning and state-of-the-art baselines, while they consume less training time and memory size.}
}
\end{center}

\subsection{RQ2: Performance in the Cross-lingual Scenario}

\begin{table}[t]
\setlength\tabcolsep{3.2pt}
\caption{Evaluation results of CMG in the cross-lingual scenario.} \label{cross-lingual table}
\centering
\begin{tabular}{cccccccc}
\hline
\multicolumn{2}{c}{Models} & Java & C\# & C++ & Python & JavaScript & Avg. \\
\hline
\multirow{3}{*}{\textbf{\thead{Java}}} & CodeT5 & - & 9.94 & 10.06 & 10.95 & 12.10 & 10.76 \\
& CodeT5-L & - & \textbf{10.52} & \textbf{11.00} & \textbf{11.70} & \textbf{13.30} & \textbf{11.63} \\
& CodeT5-A & - & 10.47 & 10.69 & 11.31 & 12.90 & 11.34 \\
\hline
\multirow{3}{*}{\textbf{\thead{C\#}}} & CodeT5 & 10.23 & - & 9.97 & 10.68 & 13.18 & 11.02 \\
& CodeT5-L & \textbf{11.01} & - & \textbf{11.04} & \textbf{11.57} & 13.21 & \textbf{11.71} \\
& CodeT5-A & 10.65 & - & 10.89 & 11.44 & \textbf{13.34} & 11.58 \\
\hline
\multirow{3}{*}{\textbf{\thead{C++}}} & CodeT5 & 10.64 & 10.42 & - & 12.08 & 13.44 & 11.65 \\
& CodeT5-L & \textbf{11.26} & \textbf{10.93} & - & \textbf{12.64} & \textbf{14.20} & \textbf{12.26} \\
& CodeT5-A & 11.00 & 10.31 & - & 12.34 & 13.52 & 11.79 \\
\hline
\multirow{3}{*}{\textbf{\thead{Python}}} & CodeT5 & 9.78 & 10.14 & 10.47 & - & 13.24 & 10.91 \\
& CodeT5-L & \textbf{10.20} & 10.54 & 10.99 & - & \textbf{13.79} & \textbf{11.38} \\
& CodeT5-A & 10.00 & \textbf{10.55} & \textbf{11.03} & - & 13.69 & 11.32 \\
\hline
\multirow{3}{*}{\textbf{\thead{JavaScript}}} & CodeT5 & 9.62 & 9.90 & 9.75 & 10.93 & - & 10.05 \\
& CodeT5-L & \textbf{11.37} & \textbf{10.73} & \textbf{11.20} & \textbf{12.30} & - & \textbf{11.40} \\
& CodeT5-A & 10.65 & 9.96 & 10.63 & 11.88 & - & 10.78 \\
\hline
\end{tabular}
\label{datasets}
\vspace{-0.5cm}
\end{table}

In this section, we investigate the performance of adapter tuning and LoRA in the cross-lingual scenario, comparing them with FMFT as well. We apply the three methods to the CMG task as its dataset contains five PLs. In the cross-lingual scenario, PLMs are expected to generate commit messages about the target PL but are fine-tuned in a different source PL. It can be seen as a combination task of monolingual CMG and program translation \cite{wang2022survey}, and is useful when the training data is limited or developers are only familiar with specific PLs. Table \ref{cross-lingual table} illustrates the evaluation results, where rows represent the source PLs while columns represent the target PLs. We only report the BLEU metric due to the page limitation.

From Table \ref{cross-lingual table}, eliminating the diagonal values, which represent source PL and target PL are the same, we could observe consistent improvements brought by adapter tuning and LoRA in the cross-lingual scenario. When comparing with the performance of FMFT, the average promotion of adapter tuning and LoRA can be up to 4.62\% and 7.56\%, respectively. It indicates that FMFT learns much PL-specific knowledge to pursue high performance, while adapter tuning and LoRA adjusting a small portion of parameters learn general-purpose knowledge, which can be transferred in diverse PLs. In addition, LoRA outperforms adapter tuning in most cross-lingual scenarios. Considering their parameter size, the result inspires us that tuning very few parameters can keep a balance between good performance and generality among different PLs, which is the superiority of adapter tuning and LoRA.

\begin{center}
\fcolorbox{black}{gray!10} 
{
\parbox{.9\linewidth}
{\textbf{Finding 3}: Adapter tuning and LoRA obtain consistent improvements than FMFT in the cross-lingual scenario. It shows their superiority in balancing PLMs' high performance and generality among different PLs.}
}
\end{center}

\subsection{RQ3: Performance in the Low-resource Scenario}

\begin{table*}[t]
\setlength\tabcolsep{2.9pt}
\caption{Evaluation results of CMG in the low-resource scenario. M is short for Meteor. R is short for Rouge-L. CodeT5, CodeT5-L, CodeT5-A represent they utilize FMFT, Adapter Tuning, and LoRA, respectively.} \label{low-resource table}
\centering
\scalebox{1}{
    \begin{tabular}{cccccccccccccccccccc}
    \hline
    \multirow{2}{*}{\textbf{Samples}} & \multirow{2}{*}{\textbf{Models}} & \multicolumn{3}{c}{\textbf{Java}} & \multicolumn{3}{c}{\textbf{C\#}} & \multicolumn{3}{c}{\textbf{C++}} & \multicolumn{3}{c}{\textbf{Python}} & \multicolumn{3}{c}{\textbf{JavaScript}} & \multicolumn{3}{c}{\textbf{Avg.}} \\
    \cline{3-20}
    & & BLEU & M & R & BLEU & M & R & BLEU & M & R & BLEU & M & R & BLEU & M & R & BLEU & M & R \\
    \hline
    \multirow{3}{*}{\textbf{\thead{1,000}}} & CodeT5 & 9.68 & \textbf{7.21} & \textbf{13.91} & \textbf{9.34} & \textbf{6.04} & \textbf{12.72} & 7.59 & 5.73 & 10.54 & \textbf{9.72} & 6.92 & 12.66 & \textbf{13.25} & \textbf{8.66} & \textbf{17.88} & \textbf{9.92} & \textbf{6.91} & \textbf{13.54} \\
    & CodeT5-L & 8.89 & 6.30 & 12.50 & 7.78 & 4.91 & 10.45 & \textbf{8.85} & \textbf{6.43} & \textbf{11.24} & 9.40 & \textbf{7.15} & \textbf{12.99} & 11.12 & 8.22 & 16.35 & 9.21 & 6.60 & 12.71 \\
    & CodeT5-A & \textbf{9.71} & 6.33 & 13.21 & 7.99 & 4.64 & 10.93 & 8.18 & 5.92 & 11.23 & 8.45 & 6.33 & 11.63 & 10.76 & 7.72 & 15.36 & 9.02 & 6.19 & 12.47 \\
    \hline
    \multirow{3}{*}{\textbf{\thead{5,000}}} & CodeT5 & \textbf{14.95} & 9.48 & \textbf{20.09} & \textbf{13.53} & \textbf{9.15} & \textbf{19.14} & 11.71 & \textbf{9.52} & \textbf{17.21} & 12.37 & 9.03 & 17.29 & 14.96 & 10.43 & 20.15 & \textbf{13.50} & \textbf{9.52} & \textbf{18.78} \\
    & CodeT5-L & 13.96 & 9.46 & 18.96 & 12.22 & 7.56 & 16.55 & 11.50 & 8.66 & 15.74 & \textbf{12.71} & \textbf{9.85} & \textbf{17.95} & 14.94 & \textbf{11.03} & \textbf{21.11} & 13.07 & 9.31 & 18.06 \\
    & CodeT5-A & 14.55 & \textbf{10.18} & 19.91 & 12.55 & 7.85 & 17.14 & \textbf{11.84} & 8.71 & 15.68 & 12.59 & 9.37 & 17.28 & \textbf{15.05} & 10.71 & 20.49 & 13.32 & 9.36 & 18.10 \\
    \hline
    \multirow{3}{*}{\textbf{\thead{10,000}}} & CodeT5 & 16.07 & 10.35 & 21.47 & 13.00 & 8.11 & 17.36 & \textbf{13.38} & \textbf{9.86} & \textbf{18.19} & 13.49 & 10.19 & 18.73 & 17.88 & 12.32 & 24.18 & 14.76 & 10.17 & 19.99 \\
    & CodeT5-L & 16.13 & 10.61 & 21.62 & 13.76 & 9.05 & 18.73 & 12.77 & 9.45 & 17.15 & 13.81 & 10.70 & 19.85 & 17.50 & 12.63 & 24.06 & 14.79 & 10.49 & 20.28 \\
    & CodeT5-A & \textbf{17.46} & \textbf{11.38} & \textbf{23.41} & \textbf{14.37} & \textbf{9.64} & \textbf{19.61} & 12.93 & 9.82 & 18.02 & \textbf{14.73} & \textbf{11.13} & \textbf{21.28} & \textbf{18.51} & \textbf{12.88} & \textbf{25.03} & \textbf{15.60} & \textbf{10.97} & \textbf{21.47} \\
    \hline
    \end{tabular}
}
\label{datasets}
\vspace{-0.5cm}
\end{table*}

% Low resource is also a common scenario for code intelligence tasks \cite{sun2022importance}. On the one hand, high-quality large-scale datasets are rare. On the other hand, training with fewer instances results in less training time. Both of these reasons encourage researchers to study the scenario persistently. In this section, we investigate how well adapter tuning and LoRA can perform in the low-resource scenario. We compare their performance with FMFT in the same sample size setting, and also analyze the difference with diverse settings. Considering that the CMG task contains hundreds of thousands of data, we randomly select 1,000, 5,000, and 10,000 training data for each PL to simulate the low-resource scenario. The evaluation results are shown in Table \ref{low-resource table}.

Low resource is also a common scenario for code intelligence tasks \cite{sun2022importance}. In this section, we investigate how well adapter tuning and LoRA can perform in the low-resource scenario. We randomly select 1,000, 5,000, and 10,000 training data for each PL in the CMG task, and the evaluation results are shown in Table \ref{low-resource table}.
From Table \ref{low-resource table}, when training with 10,000 samples ($<$ 1/10 of the full-data setting) for each PL, except for C++, it can be observed that adapter tuning and LoRA consistently outperform FMFT in the other four PLs. As for their average performance, adapter tuning obtains promotions of 5.45\%, 4.60\%, and 5.86\% for BLEU, Meteor, and Rouge-L, respectively. Accordingly, LoRA surpasses FMFT by 0.20\%, 3.17\%, and 1.48\% in each of the metrics, respectively. This demonstrates that PEFT techniques make PLMs adapt to code-change-related tasks more efficiently compared with FMFT.
Nonetheless, when reducing the training samples to 1,000 and 5,000 for each PL ($<$ 1/100 and $<$ 1/50 of the full-data setting), it can be observed that the average performance of FMFT becomes slightly better than adapter tuning and LoRA. It is also noted that, in the setting of 1,000 training samples, adapter tuning and LoRA outperform FMFT in C++, and in the setting of 5,000 training samples, adapter tuning and LoRA surpass FMFT in Python and JavaScript. Compared with the full-data setting where FMFT always outperforms PEFT techniques slightly (refer to RQ1), PEFT techniques carry their superiority in the low-resource setting. We speculate that CMG is a tough code-change-related task that needs more parameter tuning for adaptation. Even though PEFT inherently carries superiority in the low-resource setting, they still cannot exhibit a better understanding of code changes with very limited training samples in such generative tasks. It reminds us that PEFT needs a certain amount of data to learn such tricky code-change-related semantics. However, as the performance differences between PEFT techniques and FMFT are still comparable, we still suggest using PEFT techniques in the low-resource setting, as their low computational overhead. 

% However, due to the fewer parameters that adapter tuning and LoRA have, they can not surpass the performance of FMFT when training data is very limited. It can also be concluded that when increasing the sample size in the low-resource scenario, adapter tuning and LoRA can rapidly tune PLMs to downstream tasks than FMFT.

\begin{center}
\fcolorbox{black}{gray!10} 
{
\parbox{.9\linewidth}
{\textbf{Finding 4}: Adapter tuning and LoRA show effectiveness in the low-resource scenario. Practitioners are recommended to use PEFT techniques in the low-resource setting as their performance is outperforming or at least comparable to FMFT.}
}
\end{center}

\subsection{RQ4: Probing Tasks}

\begin{table}[t]
\setlength\tabcolsep{6pt}
\caption{Evaluation results of probing tasks for JIT-DP. The best average performance is highlighted in boldface.} \label{probing table}
\centering
\begin{tabular}{ccccc}
\hline
\multicolumn{2}{c}{Models} & TYP & CCM & LTP \\
\hline
\multirow{6}{*}{\textbf{\thead{FMFT}}} & CodeBERT & 54.0 & 50.0 & 62.5 \\
& GraphCodeBERT & 78.5 & 58.0 & 71.5 \\
& PLBART & 80.0 & 62.0 & 73.5 \\
& UniXcoder & 89.0 & 52.5 & 65.5 \\
& CodeT5 & 91.5 & 70.0 & 79.5 \\
\cline{2-5}
& Avg. & 78.6 & 58.5 & 70.5 \\
\hline
\multirow{6}{*}{\textbf{\thead{LoRA}}} & CodeBERT & 91.0 & 50.5 & 72.5 \\
& GraphCodeBERT & 97.5 & 61.5 & 71.5 \\
& PLBART & 87.5 & 55.5 & 73.5 \\
& UniXcoder & 95.5 & 58.5 & 67.5 \\
& CodeT5 & 92.5 & 64.5 & 77.0 \\
\cline{2-5}
& Avg. & \textbf{92.8} & 58.1 & 72.4 \\
\hline
\multirow{6}{*}{\textbf{\thead{Adapter\\Tuning}}} & CodeBERT & 91.0 & 56.0 & 65.0 \\
& GraphCodeBERT & 94.0 & 61.0 & 71.0 \\
& PLBART & 84.5 & 63.0 & 76.5 \\
& UniXcoder & 93.5 & 55.5 & 70.5 \\
& CodeT5 & 95.5 & 72.0 & 80.5 \\
\cline{2-5}
& Avg. & 91.7 & \textbf{61.5} & \textbf{72.7} \\
\hline
\end{tabular}
\label{datasets}
\vspace{-0.5cm}
\end{table}

% As shown in Table \ref{JIT-DP main}, adapter tuning and LoRA achieve prominent improvements, especially in the setting that incorporates extra expert features. 
In this section, we utilize three probing tasks, namely TYP, CCM, and LTP tasks, to explore what code properties are encoded in PLMs, thereby making an explanation for the above experimental results. TYP task relates to static semantic information, while CCM and LTP tasks are associated with dynamic semantic information from global and local perspectives. To be specific, in the JIT-DP task, we reuse the five PLMs that are trained with expert features by FMFT, adapter tuning, and LoRA, respectively. We also probe the fine-tuned PLM (i.e., CodeT5) in the CMG task among different PLs. All PLMs used here are trained in the full-data setting. Taking probing tasks' data as inputs, we extract the encoded features from the last hidden layer, which is the 6-\textit{th} layer for PLBART and the 12-\textit{th} layer for the others. Then extracted features are fed into a simple linear classifier to predict their categories \cite{karmakar2021pre}. The classifier has no hidden units, so the performance of probing tasks prohibitively depends on the feature vectors. The evaluation results of probing tasks for JIT-DP and CMG are illustrated in Table \ref{probing table} and Table \ref{probing table cmg}.

\begin{figure*}[t]
	\centering
        \setlength{\abovecaptionskip}{0.cm}
		\includegraphics[width=\textwidth]{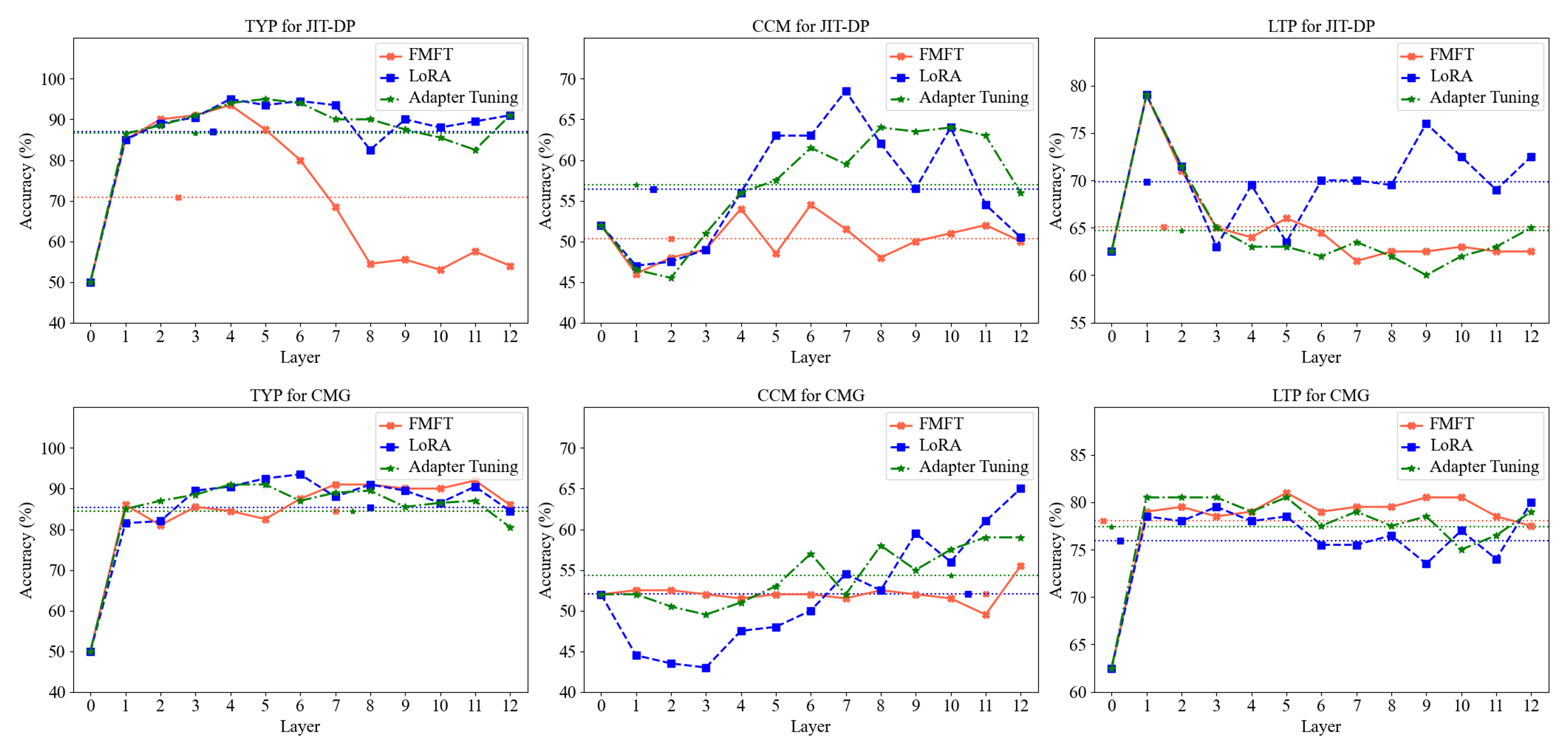}
	  \caption{Accuracy of probing tasks by layers.}\label{probe figure}
\vspace{-0.5cm}
\end{figure*}

From Table \ref{probing table}, we could derive several insightful findings: \text{(i)} The average performance of adapter tuning outperforms FMFT in all three probing tasks, while the average performance of LoRA surpasses FMFT in TYP and LTP, but degrades slightly in CCM. \text{(ii)} Comparing with the improvements obtained by adapter tuning and LoRA in the TYP task, the promotions on CCM and LTP are not dramatic. The first finding is expected and explains that the adapter tuning and LoRA learn more about code-change-related semantics from both global and local perspectives while keeping the original understanding ability of static code semantics. The second finding indicates that learning dynamic code semantics is more difficult than static semantics. Considering that code-change-related tasks are more complex, the result is reasonable. In addition, we also extract encoded features from each layer of CodeBERT to probe the layer-wise performance, as shown in the first row of Figure \ref{probe figure}. The dashed lines indicate the average performance of all layers. Except for the adapter tuning in the LTP task, consistent substantial improvements are obtained by adapter tuning and LoRA, which also shows their superiority in representing both static and dynamic code semantics.

As for the CMG task, Table \ref{probing table cmg} shows that adapter tuning and LoRA can not surpass FMFT in all three probing tasks on average of diverse PLs, but their differences are not evident. Subsequently, we further select Java to dig deeper into the layer-wise probing, as shown in the second row of Figure \ref{probe figure}. We found that no one approach can dominate all three probing tasks. Instead, they achieve an almost neck-to-neck performance on average of different layers. For example, FMFT and PEFT techniques perform almost the same in the TYP task. Adapter tuning outperforms the other two approaches in the CCM task while FMFT outperforms them in the LTP task. Hence, it is hard to say which approach makes PLMs learn more about static or dynamic code semantics. To some extent, it is also consistent with the experimental results of RQ1, where the CMG ability learned by three techniques is comparable.
% It is interesting that even though AT and LoRA can understand the code-change semantics as well as FMFT, as demonstrated by the above classification probing tasks, they still perform a little weakly in CMG tasks as shown in Table \ref{CMG main}. We speculate that CMG, as a generative task, is relatively more challenging compared with binary classification probing tasks. Tuning very few parameters indeed makes PLMs understand the code-change semantics, but it is still hard to generate the correct commit messages compared with full-model fine-tuning on the full-data setting.

\begin{center}
\fcolorbox{black}{gray!10} 
{
\parbox{.9\linewidth}
{\textbf{Finding 5}: Probing tasks show that adapter tuning and LoRA can benefit the understanding of dynamic code semantics from both the global and local perspectives in the JIT-DP task. In the CMG task, probing tasks show the differences among FMFT, adapter tuning, and LoRA are not evident. }
}
\end{center}

\section{Implications}

\begin{table}[t]
\setlength\tabcolsep{6pt}
\caption{Evaluation results of probing tasks for CMG. The best average performance is highlighted in boldface.} \label{probing table cmg}
\centering
\begin{tabular}{ccccc}
\hline
\multicolumn{2}{c}{Models} & TYP & CCM & LTP \\
\hline
\multirow{6}{*}{\textbf{\thead{FMFT}}} & Java & 86.0 & 55.5 & 77.5 \\
& C\# & 88.0 & 57.0 & 78.0 \\
& C++ & 85.5 & 57.5 & 79.0 \\
& Python & 86.0 & 51.0 & 75.5 \\
& JavaScript & 88.5 & 54.0 & 79.0 \\
\cline{2-5}
& Avg. & 86.8 & 55.0 & 77.8 \\
\hline
\multirow{6}{*}{\textbf{\thead{LoRA}}} & Java & 84.5 & 65.0 & 80.0 \\
& C\# & 91.0 & 54.5 & 77.5 \\
& C++ & 90.0 & 58.0 & 77.5 \\
& Python & 88.5 & 58.5 & 76.5 \\
& JavaScript & 86.5 & 60.5 & 73.5 \\
\cline{2-5}
& Avg. & \textbf{88.1} & 59.3 & 77.0 \\
\hline
\multirow{6}{*}{\textbf{\thead{Adapter\\Tuning}}} & Java & 80.5 & 59.0 & 79.0 \\
& C\# & 83.5 & 62.0 & 83.5 \\
& C++ & 84.0 & 68.5 & 81.0 \\
& Python & 81.5 & 65.0 & 83.5 \\
& JavaScript & 92.0 & 63.0 & 84.0 \\
\cline{2-5}
& Avg. & 84.3 & \textbf{63.5} & \textbf{82.2} \\
\hline
\end{tabular}
\label{datasets}
\vspace{-0.5cm}
\end{table}

This paper presents the first empirical study that investigates the performance of PEFT on code-change-related tasks. In this section, we discuss some implications of this work from the perspectives of developers and researchers.

\textbf{Implications for developers}. This study indicates that PEFT can outperform FMFT on the JIT-DP task, and can achieve comparable results on the CMG task with less training time and memory consumption. The results enlighten developers can leverage PEFT to replace the original FMFT methods in such code-change-related classification tasks, and for generation tasks, when the computational resources are limited, PEFT can be an effective alternative. In addition, the superiority that PEFT shows in the cross-lingual and low-resource scenarios indicates that PEFT can be used in practice when meeting the data scarcity problem. Different from previous studies that show consistent improvements brought by PEFT in code search, code clone, and code summarization tasks \cite{saberi2023utilization, wang2023one}, our work provides guidance about when and how to apply PEFT to code-change-related tasks.

\textbf{Implications for researchers}. As the first empirical study to explore the performance of PEFT in code-change-related tasks, this work obtains several interesting findings. It can be seen that in the JIT-DP task, the promotion of PEFT increases when adding the expert features, which demonstrates that PEFT can facilitate encoding the extra features into PLMs. Because expert features are more straightforward to code-change-related semantics, adjusting few parameters can transfer them to specific downstream tasks. It deserves to be investigated in other tasks to pursue high performances. We also call for more effective probing tasks designed for code-change-related tasks, probing information like syntax and structure. Besides, our study demonstrates that PEFT can be a powerful alternative approach, with their standard architectures. It also represents the need to further adapt PEFT with more code-change-related designs.

\section{Threats to Validity}

We identify the following threats to our study: 
\textbf{1) Evaluation Tasks}: We conduct experiments on two widely-studied code-change-related tasks, including a classification task, JIT-DP, and a generation task, CMG. Although they are representative, there are other code-change-related tasks like bug fixing patch identification \cite{hoang2019patchnet} and automated patch correctness assessment \cite{lin2022context}. Therefore, in the future, we plan to assess our findings on more tasks.
\textbf{2) Evaluation Datasets}: The quality of datasets can affect the evaluation performance. Thus, to mitigate the issue, we select widely-used large-scale datasets for JIT-DP and CMG. However, it is noted that there are also other datasets constructed from different source data. We will conduct experiments on more datasets to confirm our findings in future works. \textbf{3) Pre-trained Language Models}: Due to the computational resource constraints, the maximal PLM we use in our study is CodeT5, which contains 220M parameters. In the future, larger PLMs deserve to be explored. \textbf{4) Hyperparameter Setting}: In this study, we implement the standard architecture of AT and LoRA and follow their default hyperparameter settings. We also modify their intermediate dimensions, which does not bring expected improvement. Finding optimal hyperparameters is always a difficult task that is needed to explore persistently.

\section{Related Work}

\subsection{Code Change Learning}

Code changes are closely associated with software development and exist extensively. It appears when adding new features, fixing bugs, or refactoring current code \cite{yang2023significance, impactofchanges}. Along with code changes, developers have to undertake considerable workloads for tasks like writing high-quality commit messages \cite{jiang2017automatically} and identifying whether software changes are defect-inducing \cite{kamei2012large}. Therefore, techniques aiming at automating such code-change-related tasks come out and show their effectiveness in enhancing development efficiency. Most of the works follow the representation learning approach, which converts code changes into feature vectors primarily, and then retrieves in the feature space to obtain expected results.

Concentrating on concrete downstream tasks, some studies propose task-specific methods to learn code change representations. For commit message generation, CoDiSum \cite{xu2019commit} extracts code structure and code semantics by separate bidirectional GRU and aligns the two parts by an attention layer. FIRA \cite{dong2022fira} describes code change operations in fine-grained graphs and utilizes a graph-neural-network-based encoder to learn representations. RACE \cite{shi-etal-2022-race}, based on Transformer architecture, retrieves instructive code changes firstly by the cosine similarity of representation vectors, using them to guide the generation of commit messages. For just-in-time defect prediction, DeepJIT \cite{hoang2019deepjit} leverages two discrete CNNs to extract features from code changes and corresponding commit messages, and employs a fully connected network for feature fusion. JITLine \cite{pornprasit2021jitline} extracts bag-of-tokens features and conducts five well-known classification techniques, like Support Vector Machine (SVM), to build commit-level prediction models. JIT-Fine \cite{ni2022best} combines 14 change-level expert features \cite{kamei2012large} with extracted semantic features, obtaining integrated representations. 

There are also some works focusing on general-purpose code change representations. CC2Vec \cite{hoang2020cc2vec} inputs the removed code and added code separately to a hierarchical attention network, extracts their features, and uses multiple comparison functions to fuse the two parts. CCRep \cite{liu2023ccrep} proposes a novel mechanism called query back, which can emphasize the core status of changed code and adaptively learn representations from the context. CodeReviewer pre-trains PLMs based on four proposed pre-training tasks in the code review scenario, which makes CodeReviewer \cite{li2022automating} better represent code changes in downstream tasks. Similarly, CCT5 \cite{lin2023cct5} is designed for code-change-related tasks. It proposes five pre-training tasks to build the semantic connection between the changed codes and corresponding commit messages. In this paper, focusing on the performance that PEFT methods can achieve in code-change-related tasks, we utilize PEFT methods instead of the original FMFT to learn dynamic code semantics.

\subsection{Pre-trained Language Models of Code}

Inspired by the effectiveness and versatility of PLMs shown in the NLP field, a range of PLMs of code that take into account code-specific characteristics arise. Although they all based on Transformer architecture, PLMs of code can be subdivided into three categories: Encoder-only, Decoder-only, and Encoder-Decoder \cite{niu2023empirical, zeng2022extensive}. CodeBERT \cite{feng-etal-2020-codebert} and GraphCodeBERT \cite{DBLP:conf/iclr/GuoRLFT0ZDSFTDC21} are two representative encoder-only models. They are both based on BERT \cite{DBLP:conf/naacl/DevlinCLT19} and are pre-trained with natural and programming languages. In addition, GraphCodeBERT introduces semantic-level structural information by adding data flow in the pre-training stage. Decoder-only models of code, like GPT-C \cite{svya2020gptc}, CodeGen \cite{nijkamp2022codegen}, and Code Llama \cite{roziere2023code}, are on the basis of GPT series \cite{radford2019language}, which predicts text when given the preceding context. Some recent typical encoder-decoder models are PLBART \cite{ahmad-etal-2021-unified}, UniXcoder \cite{guo-etal-2022-unixcoder}, and CodeT5 \cite{wang-etal-2021-codet5}. PLBART uses the same architecture as BART \cite{lewis-etal-2020-bart}, and is pre-trained via denoising autoencoding. UniXcoder leverages cross-modal contents like ASTs and code comments to enrich its pre-training corpus, and can be better used for auto-regressive tasks. CodeT5 is based on T5 \cite{JMLR:v21:20-074} architecture and proposes to preferably capture code semantics via developer-assigned identifiers. In this paper, we apply PEFT methods to a range of PLMs to investigate their performances and generalization abilities.

\section{Conclusion}

For code-change-related tasks, there is an obvious gap between the pre-training and the fine-tuning processes, compared to static code comprehension. Thus, whether PEFT can outperform FMFT is still an open question. In this paper, we experimentally investigate the performance of two prevalent PEFT methods, namely AT and LoRA, on two widely-studied code-change-related tasks, including JIT-DP and CMG. Our study shows that AT and LoRA can achieve the SOTA results on JIT-DP and comparable performances on CMG with less training time and memory consumption, which indicates their effectiveness and efficiency. Even in the cross-lingual and low-resource scenarios, they also exhibit superiority. To make an explanation for the performance of AT, LoRA, and FMFT, we also conduct three probing tasks to measure their code semantic learning from both static and dynamic perspectives.
% Probing tasks certify that PEFT can be beneficial to the comprehension of dynamic code semantics from both global and local perspectives. 
Finally, we summarize our findings and provide implications to enlighten future works.

% This study addresses the existing gap between the pre-training and fine-tuning processes in code-change-related tasks and investigates the potential of Parameter Efficient Fine-Tuning (PEFT) methods to outperform Full-Model Fine-Tuning (FMFT). Specifically, we experimentally evaluate two prevalent PEFT methods, namely Adapter Tuning (AT) and Low-Rank Adaptation (LoRA), on two widely-studied code-change-related tasks: Just-In-Time Defect Prediction (JIT-DP) and Commit Message Generation (CMG).

% Our experimental results demonstrate that both AT and LoRA achieve state-of-the-art (SOTA) performance on JIT-DP and comparable results on CMG when compared to FMFT and other approaches. Importantly, these methods achieve these results with considerably reduced training time and memory consumption, highlighting their effectiveness and efficiency. Furthermore, even in cross-lingual and low-resource scenarios, AT and LoRA exhibit superiority.

% Overall, our findings contribute to a better understanding of the potential of PEFT methods in code-change-related tasks. The superior performance, efficiency, and cross-lingual adaptability of AT and LoRA highlight their practical utility. These results provide implications for future research in the field of code change learning. Further exploration and development of PEFT methods can enhance the performance and efficiency of adapting pre-trained language models to code-change-related tasks, ultimately advancing the capabilities of code comprehension systems.
\section*{Acknowledgment}

This work is supported in part by the General Research Fund (GRF) of the Research Grants Council of Hong Kong, and the industry research funds of City University of Hong Kong (7005217,9220097,9220103,9229029,9229098,9678149), also by the National Natural Science Foundation of China under Grant No. 62302021.

% % Loading bibliography database
\bibliographystyle{plain}
\bibliography{myrefs}

% \vspace{12pt}
% \color{red}
% IEEE conference templates contain guidance text for composing and formatting conference papers. Please ensure that all template text is removed from your conference paper prior to submission to the conference. Failure to remove the template text from your paper may result in your paper not being published.

\end{document}